\documentclass[iop]{emulateapj}

\newcommand{\kpc}{\mbox{$\,h^{-1}$kpc}}
\newcommand{\mpc}{\mbox{$\,h^{-1}$Mpc}}
\newcommand{\gpc}{\mbox{$\,h^{-1}$Gpc}}
\newcommand{\mpci}{\mbox{$\,h$\,Mpc$^{-1}$}}
\newcommand{\Msun}{\mbox{$\,h^{-1}$M$_\sun$}}

\begin{document}

\title{Mass-Dependent Baryon Acoustic Oscillation Signal and 
Halo Bias}

\author{Qiao Wang and Hu Zhan}
\affil{Key Laboratory of Optical Astronomy,
National Astronomical Observatories, 
Chinese Academy of Sciences, Beijing 100012, China}
\email{qwang@nao.cas.cn}

\begin{abstract}
We characterize the baryon acoustic oscillations (BAO) feature in 
halo two-point statistics using \emph{N}-body simulations. We find 
that nonlinear damping of the BAO signal is less severe for halos in the mass range we investigate
than for dark matter. The amount of damping depends weakly on the 
halo mass. The correlation functions show a mass-dependent drop 
of the halo clustering bias below roughly $90\mpc$, which coincides
with the scale of the BAO trough. The drop of bias 
is 4\% for halos with mass $M>10^{14}\Msun$ and reduces to 
roughly 2\% for halos with mass $M>10^{13}\Msun$. In contrast, halo
biases in simulations without BAO change more smoothly around 
$90\mpc$. In Fourier space, 
the bias of $M>10^{14}\Msun$ halos decreases smoothly by $11\%$ from 
wavenumber $k = 0.012\mpci$ to $0.2\mpci$, whereas that of 
$M>10^{13}\Msun$ halos decreases by less than $4\%$ over the 
same range. By comparing the halo biases in pairs of otherwise 
identical simulations, one with and the other without BAO, 
we also observe a modulation of the halo bias. 
These results suggest that 
precise calibrations of the mass-dependent BAO signal and 
scale-dependent bias on large scales would be needed for 
interpreting precise measurements of the two-point statistics of 
clusters or massive galaxies in the future.

\end{abstract}
\keywords{Cosmology: theory --- large-scale structure of Universe --- 
dark matter}

\section{Introduction}

Baryon Acoustic Oscillations \citep[BAO,][]{peebles1970, sunyaev1970}
produce an imprint on 
the matter distribution by the perturbations in the photon-baryon 
fluid before recombination. It provides a standard ruler for probing 
dark energy \citep[e.g.,][]{eisenstein1998,cooray2001,blake2003,hu2003,
linder2003,seo2003},
and is considered to be least affected by systematics 
\citep{DETF2006}. Thus far, the BAO signal has been detected in samples
from Sloan Digital Sky Survey \citep{eisenstein2005, padmanabhan2007, percival2010}, 
2dF Galaxy Redshift Survey \citep{cole2005}, 
6dF Galaxy Survey \citep{beutler2011},
WiggleZ Dark Energy Survey \citep{blake2011}, and 
Baryon Oscillation Spectroscopic Survey \citep{anderson2012}. 
Future spectroscopic and imaging surveys, such as BigBOSS, Euclid, 
and Large Synoptic Survey Telescope, all aim to determine cosmic 
distances to sub-percent level precision with the BAO technique 
\citep[e.g.,][]{zhan2009, BIGBOSS2011}.

Although the BAO signal does not evolve in linear theory, nonlinearity
causes a slight shift of its scale and significant damping 
of its amplitude \citep[e.g.,][]{seo2008,seo2010,sherwin2012,mccullagh2013}. These effects must be 
modeled or corrected in BAO analyses. Given the 
unprecedented statistical power of future surveys, it is necessary
to calibrate the nonlinear BAO signal accurately. Efforts have been 
made for dark matter with both \emph{N}-body simulations 
\citep{smith2008,padmanabham2009} and perturbation theory 
\citep{jeong2006,crocce2008, matsubara2008}. 

It is reasonable to assume that the halo bias is constant
on large scales according to the ``peak-background split'' or 
``local bias'' model\citep[e.g.,][]{fry1993, mo1996, scherrer1998}.
For precision cosmology, however, one should quantify the scales 
of validity as well as any departure from constancy
for halos of different masses. Investigations of the halo BAO signal
have given hints for scale-dependence in the halo bias even at the 
BAO scale \citep[e.g.,][]{seo2005, noh2009, desjacques2010}. 
In this work, we determine the halo correlation functions and power 
spectra with high precision and clearly 
demonstrate the mass-dependent halo BAO signal and behavior of 
the halo bias beyond the BAO scale.

\section{Simulations\label{sec:simu}}

We run two sets of \emph{N}-body simulations to study the halo BAO 
signal and halo bias at redshift $z=0$. The initial matter power 
spectrum of the first set 
is calculated using {\sc camb} \citep*{CAMB} with the parameters: 
$\Omega_{\Lambda}$=0.73, $\Omega_\mathrm{M} = 0.27$, $h = 0.71$, 
$\sigma_8 = 0.8$, $n_s = 1$, and $\Omega_\mathrm{b}=0.045$.
That of the second set is calculated using the fitting formula 
in \citet{eisenstein1998} with the same parameters but no BAO 
wiggles. For convenience, we refer to them as the BAO 
simulations and the no-BAO simulations, respectively.

The simulation box is $1 \gpc$ on each side. With $N_\mathrm{p}=640^3$ 
particles, the mass resolution is $2.86 \times 10^{11}\Msun$, and the 
softening length is set to $50 \kpc$. 
The simulations are run with {\sc gadget}-2 \citep{GADGET2} from 
initial redshift $z_i=100$ to $z=0$.

There are 150 realizations in each set of the simulations. The 
initial conditions are generated using 2LPTic \citep{2LPTic} with
the same random seed for a pair of simulations, one in each set. 
In this way, the effect of sample variance can be 
reduced in the comparison between the results from the two 
sets of simulations.

Halos are identified using the Friends-of-Friends group finder 
from University of Washington's \emph{N}-Body 
Shop\footnote{http://www-hpcc.astro.washington.edu/}.
The linking length is set to 0.2 times the mean 
particle separation. Given the mass resolution, we 
limit our study to halos more massive than $10^{13}\Msun$, which
corresponds to about 35 simulation particles. 
The obtained comoving halo number density at $z=0$ is 
$4.8 \times 10^{-4}h^3\mbox{Mpc}^{-3}$ for $M> 10^{13}\Msun$ and 
$2.5 \times 10^{-5}h^3\mbox{Mpc}^{-3}$ for $M> 10^{14}\Msun$.

\section{Halo BAO and Halo Bias \label{sec:bias}}

\subsection{BAO in Halo Correlation Functions}

We calculate the halo two-point correlation functions $\xi^\mathrm{hh}(r)$ with 
the Landy-Szalay estimator \citep{landy1993} 
\begin{equation} \label{eq:xi_ls^hh}
\xi^\mathrm{hh}(r) = \frac{N_\mathrm{r}^2}{RR(r)}\left[\frac{HH(r)}{N_\mathrm{h}^2} - 
\frac{2HR(r)}{N_\mathrm{h}N_\mathrm{r}} + \frac{RR(r)}{N_\mathrm{r}^2}\right],
\end{equation}
where $N_\mathrm{r}$ is the size of a uniform random sample, 
$N_\mathrm{h}$ is the size of the halo sample, $HH$ is the number of 
halo pairs with a separation between $r-\Delta r/2$ and 
$r+\Delta r/2$, $HR$ is the number of pairs between the halo sample 
and the uniform random sample, and $RR$ is that within the uniform 
random sample. Since the halo sample from the simulations has 
periodic boundaries and a uniform selection function, 
Equation~(\ref{eq:xi_ls^hh}) reduces, for an infinitely dense 
random sample, to
\begin{equation} \label{eq:xi^hh-simple}
\xi^\mathrm{hh}(r)=\frac{HH(r)}{N_\mathrm{h}^2}\frac{V_\mathrm{box}}{V_\mathrm{bin}(r)} - 1,
\end{equation}
where $V_\mathrm{box}$ is the volume of the simulation box, and $V_\mathrm{bin}(r)$ 
is the volume of the spherical shell between 
$r-\Delta r/2$ and $r+\Delta r/2$.

\begin{figure}
\centering
\includegraphics[width=0.95\columnwidth]{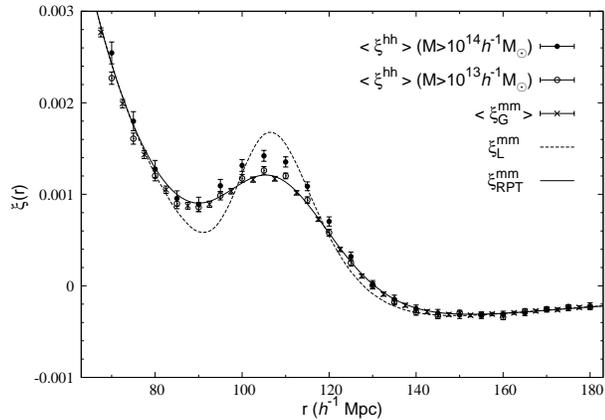}
\caption{Mass-dependent halo BAO signal at $z=0$. 
The symbols represent the correlation functions of $M > 10^{14}\Msun$ 
halos (filled circles),  $M > 10^{13}\Msun$ halos (open circles),
and dark matter (crosses) in the BAO simulations. 
The error bars show 1-$\sigma$ errors of the mean correlation 
functions of 150 realizations. 
The dashed line is the linear dark matter correlation 
function, and the solid line is the nonlinear dark matter correlation 
function from the RPT fitting formula.
These correlation functions are normalized in the range of 
$160$--$180\mpc$. 
\label{fig:massdep}}
\end{figure}

The correlation functions $\langle\xi^\mathrm{hh}\rangle$ averaged over 
the ensemble of the 150 BAO simulations are 
shown in Figure~\ref{fig:massdep} for halos of mass 
$M > 10^{13}\Msun$ (open circles) and $M > 10^{14}\Msun$ 
(filled circles). 
For comparison, we also plot the nonlinear dark matter correlation 
function from the simulations, i.e., the correlation of all 
simulation particles ($\langle\xi^\mathrm{mm}_\mathrm{G}\rangle$, 
crosses), the linear dark matter correlation function 
($\xi_\mathrm{L}^\mathrm{mm}$, dashed line), and the nonlinear dark 
matter correlation function from a convolution of 
$\xi_\mathrm{L}^\mathrm{mm}$ with the Gaussian damping factor 
predicted by Renormalized Perturbation Theory 
\citep[RPT,][]{crocce2006b, crocce2008} ($\xi_\mathrm{RPT}^\mathrm{mm}$, 
solid line). Because it is time consuming to estimate
the dark matter correlation function from $640^3$ particles with the 
pair-counting method, we assign the particles onto a density grid 
(hence the subscript G) of $200^3$ cells with the 
cloud-in-cell assignment function \citep{hockney1981} and then 
calculate the correlation function from pairs of the cells.

Studies have shown that nonlinear evolution damps the BAO feature 
\citep[e.g.,][]{eisenstein2007damp, smith2008, seo2008}. 
Such an effect on dark matter can be fairly accurately modeled with 
the RPT formulism (see, e.g., Figure~\ref{fig:massdep}). 
Comparing the correlation functions from 
the simulations, we find that the damping effect of the BAO signal 
depends on the tracer's mass, which cannot be scaled 
away by a constant clustering bias. As explained in 
\citet{eisenstein2007damp}, the BAO damping is due to the 
motions of matter or tracers relative to their initial separation,
i.e., the Lagrangian displacements between pairs.
The mass-dependent damping of the halo BAO signal suggests 
that the characteristics of halo motions are also mass-dependent.
For a test, we measure the pairwise velocity dispersion 
$\sigma_\mathrm{pv}$, which is proportional to the rms Lagrangian 
displacement in the Zel'dovich approximation, and find that 
$\sigma_\mathrm{pv} = 650.1 \pm 6.0 \mbox{km\,s$^{-1}$}$ for 
$M>10^{13}\Msun$ halos and 
$\sigma_\mathrm{pv} = 632.5 \pm 8.4 \mbox{km\,s$^{-1}$}$ for 
$M>10^{14}\Msun$ halos over the separations of $90$--$120\mpc$.
These results appear to be consistent with
Figure~\ref{fig:massdep} in the sense that larger pairwise 
random motions would damp the BAO signal more. However, we note 
that nonlinear evolution weakens the correspondence betweeen the
pairwise velocity and the Lagrangian displacement. 
A more thorough investigation is needed to fully understand the
mass-dependent BAO damping.

\subsection{Halo Bias from Correlation Functions \label{sec:bump}}

The difference between the halo correlation functions in 
Figure~\ref{fig:massdep} can be accommodated with scale-dependent 
biases. We estimate the halo bias in two ways:
\begin{eqnarray}
b_\mathrm{CC}(r)   & \equiv & \langle\xi^\mathrm{hh}(r)/\xi^\mathrm{hm}(r)\rangle, 
\label{eq:bcc} \\
b_\mathrm{G}(r) & \equiv & \langle\xi_\mathrm{G}^\mathrm{hh}(r)
\rangle^\frac{1}{2} / \langle\xi_\mathrm{G}^\mathrm{mm}(r)\rangle^\frac{1}{2}, 
\label{eq:bdc}
\end{eqnarray}
where $\xi^\mathrm{hm}$ is the cross correlation between the halos and 
dark matter in each simulation, and $\xi_\mathrm{G}^\mathrm{hh}$ is 
the correlation function calculated from the halo density grid 
(analogous to $\xi_\mathrm{G}^\mathrm{mm}$). The cross correlation 
is given by
\begin{equation} 
\xi^\mathrm{hm}(r)=\frac{HM(r)}{N_\mathrm{h} N_\mathrm{p}}\frac{V_\mathrm{box}}{V_\mathrm{bin}(r)} - 1,
\end{equation}
where $HM$ is the number of halo--simulation particle pairs with a 
separation between $r-\Delta r/2$ and $r+\Delta r/2$. One might 
attempt to take an ensemble average of the ratio
$\xi_\mathrm{G}^\mathrm{hh} / \xi_\mathrm{G}^\mathrm{mm}$ as an estimator for $b^2$. 
Such a ratio can fluctuate wildly and even be negative in 
some realizations when the correlations are close to zero.
Equation~(\ref{eq:bdc}) is better behaved, though the covariance
between $\xi_\mathrm{G}^\mathrm{hh}$ and $\xi_\mathrm{G}^\mathrm{mm}$
needs to be accounted for when determining the uncertainties of
$b_\mathrm{G}$.

\begin{figure}
\centering
\includegraphics[width=0.95\columnwidth]{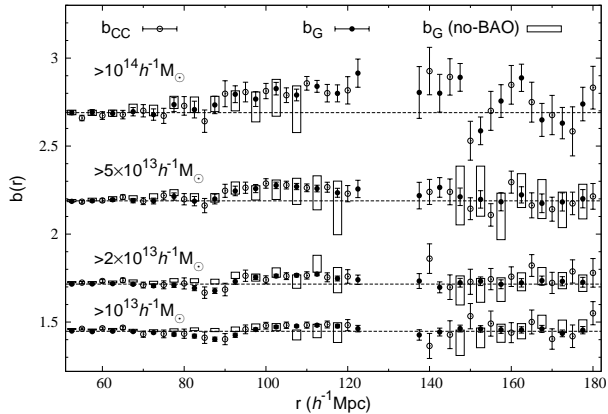}
\caption{Halo bias from the correlation functions. 
Open circles and filled circles represent, respectively, 
$b_\mathrm{CC}$ and $b_\mathrm{G}$ (see Equations~[\ref{eq:bcc}] and 
[\ref{eq:bdc}]) in the BAO simulations. 
Horizontal lines mark the 
average values of $b_\mathrm{G}$ between $50\mpc$ and $70\mpc$. 
Boxes represent $b_\mathrm{G}$ in the no-BAO simulations.
\label{fig:br}}
\end{figure}

Figure~\ref{fig:br} shows the behavior of the halo bias for four 
mass ranges as labeled. 
Since the halo mass function is fairly steep at $M > 10^{13} \Msun$, 
each mass range is dominated by halos at its low-mass 
end. The halo bias cannot be determined accurately when the 
correlation functions are close to zero. Therefore, we leave a 
gap where the errors become too large to be informative.

For the BAO simulations, the results of the two different 
estimators, $b_\mathrm{CC}$ (open circles) and $b_\mathrm{G}$ (filled 
circles), are consistent with each other over the scales and masses 
of interest. They both show a systematic offset between the 
average halo bias over $50$--$70\mpc$ (dashed lines) and that over 
the BAO peak (roughly $95$--$115\mpc$) for all the four mass 
ranges. The decrement of the bias below the transition 
scale depends on the halo mass: 4\% for 
$M > 10^{14}\Msun$ and roughly 2\% for $M > 10^{13}\Msun$. 
This finding is relevant to BAO measurements from 
luminous red galaxies, which are mostly central galaxies in halos 
of masses from a few times $10^{13}\Msun$ to $10^{14}\Msun$
\citep{zheng2009}. 

To see whether it is a coincidence that the bias drops near 
the trough of the BAO signal, we contrast the results from the
BAO simulations (circles) with those from the no-BAO simulations 
(boxes) in Figure~\ref{fig:br}. Since the latter changes
gradually around $90\mpc$, we conclude that the drop of the halo 
bias in the BAO simulations is associated with the presence of
BAO.

The scale dependence of the halo bias around the BAO scale has been 
studied previously. For example, one may find a hint for a bump of 
the halo bias around $95 \mpc$ for halos with mass 
$M \gtrsim 10^{14}\Msun$ in \citet*{manera2010} and 
\citet{manera2011}. A peak of the halo bias 
($M > 1.2\times 10^{14}\Msun$) near $100\mpc$
is also detected  in \citet{desjacques2010}. 

This work improves the statistics of the halo bias around the 
BAO scale. The mass-dependent drop of the bias below $90\mpc$ 
is well detected for halo masses down to $10^{13}\Msun$. Our results 
are consistent with a flat bias over $95$--$115\mpc$. We also 
extend the scale of interest up to $180 \mpc$ where halos are 
anticorrelated (so is dark matter). Because the correlation 
functions are close to zero in the range of $140$--$180\mpc$, 
the uncertainties of the halo bias in this range are considerably 
larger than those below $120\mpc$. Nevertheless, the results are 
consistent with a flat bias over $140$--$180\mpc$.

\subsection{Halo Bias from Power Spectra}

One can also estimate the halo bias from the halo power 
spectrum $P^\mathrm{hh}(k)$ and the dark matter power 
spectrum $P^\mathrm{mm}(k)$ with
\begin{equation}
b_\mathrm{PS}(k) \equiv \left\langle\big[P^\mathrm{hh}(k)/
P^\mathrm{mm}(k)\big]^\frac{1}{2}\right\rangle.
\end{equation}
To obtain the power spectra, we assign the halos and
simulation particles to density grids of $512^3$ cells 
using the Daubechies D12 wavelet scaling function, which
has superior performance for measuring the power 
spectrum \citep{cui2008}. We then take 
$P^\mathrm{hh}(k) = \langle|\hat{\delta}^\mathrm{h}
(\mathbf{k})|_{|\mathbf{k}|=k}^2\rangle_\mathrm{G} - 
V_\mathrm{box}N_\mathrm{h}^{-1}$ and 
$P^\mathrm{mm}(k) = \langle
|\hat{\delta}^\mathrm{m}(\mathbf{k})|_{|\mathbf{k}|=k}^2
\rangle_\mathrm{G} - V_\mathrm{box}N_\mathrm{p}^{-1}$, 
where $\hat{\delta}^\mathrm{h}(\mathbf{k})$ and 
$\hat{\delta}^\mathrm{m}(\mathbf{k})$ are the \emph{discrete} 
Fourier modes of the halo grid and the dark matter grid, 
respectively, and $\langle\ldots\rangle_\mathrm{G}$ denotes an 
average within each grid.

\begin{figure}
\centering
\includegraphics[width=0.95\columnwidth]{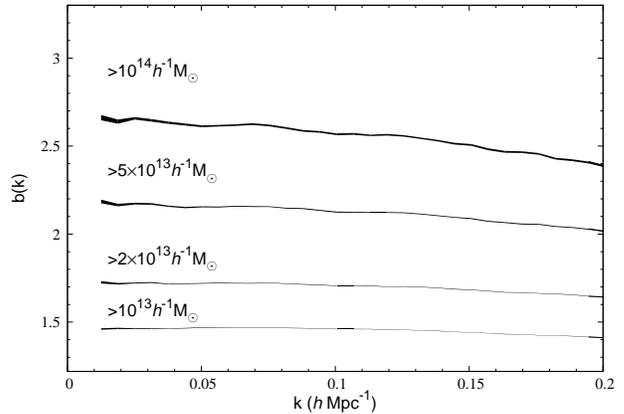}
\caption{Halo bias from the power spectra. 
The width of each curve represents 1-$\sigma$ uncertainties of 
the mean bias of the ensemble.
\label{fig:bk}}
\end{figure}

Figure~\ref{fig:bk} presents the halo bias determined from the power 
spectra for
the same mass ranges as in Figure~\ref{fig:br}. Although one does
not expect the halo bias to be absolutely constant on large scales, 
it is still surprising that the bias of $M>10^{14}\Msun$ halos 
falls by 11\% from $k=0.012 \mpci$ to $k=0.2 \mpci$. 
For halos of mass $M>10^{13}\Msun$, the bias falls by less than 
4\% over the same scales, consistent with the result of 
$M > 1.11\times 10^{13}\Msun$ in \citet*{pollack2012}. 

One may notice that there are slight oscillations in the halo bias,
which are more pronounced for more massive halos. 
This behavior is discussed in the next subsection. 

\subsection{Damping of BAO and Bias Modulation}

\begin{figure}
\centering
\includegraphics[width=0.95\columnwidth]{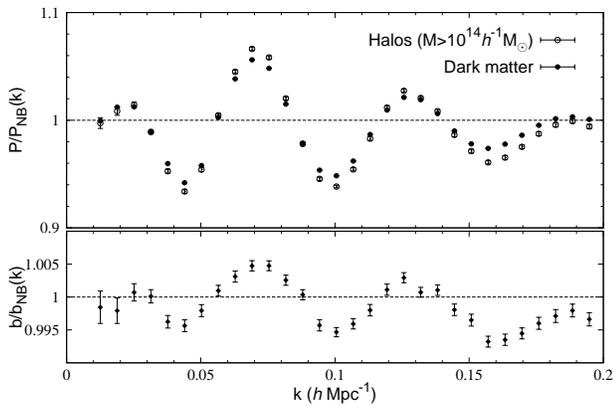}
\caption{\emph{Upper panel}: Ratio of the power spectrum from the 
BAO simulations to that from the no-BAO simulations. Filled circles 
represent dark matter, and open circles represent $M > 10^{14}\Msun$ 
halos. 
\emph{Lower panel}: Ratio of the halo bias in the BAO simulations 
to that in no-BAO simulations for $M > 10^{14}\Msun$ halos. 
\label{fig:db} }
\end{figure}

We use the no-BAO simulations as references to examine the damping 
of the BAO signal in Fourier space. The halo and dark matter power 
spectra from each simulation with BAO are divided, respectively, by 
those from the corresponding no-BAO simulation. Because 
the pair of simulations share the same random seed for their initial 
conditions, the sample variance is greatly reduced. The power spectrum
ratios are shown in the upper panel of Figure~\ref{fig:db} for 
$M > 10^{14}\Msun$ halos ($P^\mathrm{hh}/P_\mathrm{NB}^\mathrm{hh}$, open circles) and 
dark matter ($P^\mathrm{mm}/P_\mathrm{NB}^\mathrm{mm}$, filled circles). 
One can see that the BAO signal in the massive halo power spectrum suffers 
less damping than that in the dark matter power spectrum. 

There is a subtle difference between Figure~\ref{fig:massdep} and 
Figure~\ref{fig:db}. Figure~\ref{fig:massdep} illustrates a
mismatch between the shape of the halo correlation functions and 
that of dark matter around the BAO scale. It does not associate 
the very presence of BAO with the mismatch. The latter 
could be an intrinsic difference between the halos and dark matter 
that happens to occur around the BAO scale. Figure~\ref{fig:bk} 
establishes such a link, but the uncertainty is a little high for
$M > 10^{14}\Msun$ halos. The power spectrum ratios in 
Figure~\ref{fig:db} suppress the intrinsic difference between 
halos and dark matter and reduce uncertainties arising from 
the sample variance. It demonstrates for $M > 10^{14}\Msun$ halos
that the difference between the halo and dark matter BAO signals 
is truly due to the difference in the BAO damping mechanism. 

By definition, the above effect is attributed to the halo bias. 
The lower panel of Figure~\ref{fig:db} shows the ratio of the halo
bias in the BAO simulations to that in no-BAO simulations for 
$M > 10^{14}\Msun$ halos (same as the halo data divided by the dark 
matter data in the upper panel). The result oscillates in 
phase with BAO with an amplitude of $0.5\%$, suggesting a 
modulation of the halo bias by BAO. This would not be sensible 
if halos distributed exactly in proportion to dark matter. 
However, if there is nonlinearity in the halo bias, as already 
evident from its scale dependence in Figures~\ref{fig:br} and 
\ref{fig:bk}, the bias will depend on the shape of 
the dark matter power spectrum to some extent. 

\section{Discussion\label{sec:disc}}

We have demonstrated with high significance that the BAO signal 
in the halo distribution is dependent on the halo mass. In the 
correlation function, the relative amplitude of the BAO peak of 
$M>10^{14}\Msun$ halos is considerably higher than that of 
$M>10^{13}\Msun$ halos, which in turn is higher than that of dark 
matter. Since galaxies reside in halos, the BAO signal in the 
galaxy distribution must also depend on the mass of the host 
halos. 

We further show that the halo bias is scale-dependent even 
on very large scales. This could be a significant source of 
systematic errors when one extracts cosmological information 
\citep[e.g., neutrino masses,][]{lahav10}, from
the overall shape of the galaxy power spectrum.

Recently, \citet{hong2012} detect a higher-than-expected BAO 
peak in the correlation function of galaxy clusters 
($M \gtrsim 10^{14}\Msun$). Since they fix the baryon fraction 
while fitting other parameters, 
the excess BAO signal has to be compensated by a very low matter 
fraction. Although the difference between the halo and dark matter BAO 
signals in our study is not sufficient to explain all the excess 
BAO signal in \citet{hong2012}, it should be nonetheless a 
significant contributing factor.

One might expect the trend with halo mass in Figures~\ref{fig:br} 
and \ref{fig:bk} to continue below $10^{13}\Msun$. 
Investigation with $M>1.37\times 10^{12}\Msun$ halos 
\citep{sato2011} shows that the bias of such halos at $z\ge 0.5$
varies significantly with scales. A simple extrapolation of their 
results to $z=0$ suggests that the fractional variation of the 
bias of $M>1.37\times 10^{12}\Msun$ halos could be even larger 
than that of $M> 10^{13}\Msun$ halos. Therefore, it is prudent 
not to extrapolate our results to lower masses.

Besides damping the BAO signal, nonlinear evolution also shifts 
the acoustic scale slightly. We estimate this effect using 
the fitting formulae in \citet{seo2008} and find that the 
fractional shift of the BAO feature toward higher wavenumber is
$(0.28 \pm 0.16)\%$, $(0.47\pm 0.20)\%$, $(0.54\pm 0.28)\%$, and 
$(0.62\pm 0.23)\%$ for dark matter, $M>10^{13}\Msun$ halos, 
$M>2\times10^{13}\Msun$ halos, and $M>5\times10^{13}\Msun$ halos, 
respectively. These results are consistent with those in 
\citet{seo2010} and \citet{sherwin2012}. The simple fitting does 
not converge well for the highest mass range in our study. In this 
case, one can use the more sophisticated fitting formula for halos
proposed in \citet{padmanabham2009}.

Future galaxy surveys could probe effective volumes of a 
few hundred $\mbox{Gpc}^3$, similar to the total volume of 
the simulations in this study. The effect of mass-dependent 
halo BAO signal and the scale-dependent halo bias on large 
scales could be easily detected in the galaxy statistics. 
In other words, it will be necessary to accurately calibrate 
these effects for precision cosmology in the future

\acknowledgments
This work was supported by the National Natural Science foundation of 
China grant No. 11033005, the National Key Basic Research Science 
Foundation of China grant No. 2010CB833000, and the
Bairen program from the Chinese Academy of Sciences.

%\bibliographystyle{apj}
%\bibliography{ms}

\end{document}